\documentclass[12pt,preprint]{aastex}

\usepackage{amssymb}
\usepackage{graphicx}

\newcommand{\zzero}{Z$_{\odot}$}  
\newcommand{\xlf}{XLF}
\newcommand{\hst}{{\em HST}}

\newcommand{\chandra}{{\em Chandra}}

\newcommand{\startrack}{{\it StarTrack}}
\newcommand{\kev}{\hbox{{\rm\thinspace keV}}}

\newcommand{\ts}{\thinspace}

\newcommand{\eqref}[1]{(\ref{#1})}                                          
\newcommand{\sd}{\ts --\ts}                                                 

\shorttitle{\textit{Chandra} Observations of the Collisional Ring Galaxy NGC 922 }
\shortauthors{Prestwich et al.}
\bigskip

\begin{document}

\title{\textit{Chandra} Observations of the Collisional Ring Galaxy NGC 922}

\author{A.H. Prestwich}
\affil{Harvard-Smithsonian Center for Astrophysics, 60 Garden Street, Cambridge, MA 02138}
\author{J.L. Galache}
\affil{Harvard-Smithsonian Center for Astrophysics, 60 Garden Street, Cambridge, MA 02138}
\author{T. Linden}
\affil{University of California, Santa Cruz}
\author{V. Kalogera}
\affil{Northwestern University}
\author{A. Zezas}
\affil{University of Crete, Greece and Harvard-Smithsoniam Center for Astrophysics}
\author{T.P.  Roberts}
\affil{ Dept of Physics, Unversity of Durham, South Road, Durham DH1 3LE, UK}
\author{R. Kilgard}
\affil{Wesleyan University}
\author{A. Wolter}
\affil{Istituto Nazionale di Astrofisica, Milano, Italy}
\author{G. Trinchieri}
\affil{Istituto Nazionale di Astrofisica, Milano, Italy}

\begin{abstract}
In this paper we report on \chandra\ observations of the starburst galaxy  NGC~922.  NGC~922 is a drop-through ring galaxy  with an expanding ring of star formation, similar in many respects to the Cartwheel galaxy.   The Cartwheel galaxy is famous for hosting  12 ULX,  most of which are in the star forming ring.  This is the largest number of ULX seen in a single system, and has led to speculation that the low metallicity of the Cartwheel (0.3 \zzero) may optimize the conditions for ULX formation.   In contrast, NGC 922 has metallicity near solar.  The \chandra\ observations reveal a population of bright X-ray sources, including 7 ULX.   The number of ULX in NGC 922 and the Cartwheel scales with the star formation rate: we do not find any evidence for an excess of sources in the Cartwheel.   Simulations of the binary population in these galaxies suggest that the ULX population in both systems is  dominated by systems with strong wind accretion from supergiant donors onto direct-collapse BHs.   The simulations correctly predict the ratio of the number of sources in NGC 922 and the Cartwheel.   Thus it would appear that the the metallicity of the Cartwheel is not low enough to see a difference in the ULX population compared to NGC 922.   
\end{abstract}

\keywords{galaxies: individual (NGC 922) -- X-rays: galaxies}

\section{Introduction}

Chandra images of starburst galaxies are spectacular, showing a
multitude of bright point sources and plumes of X-ray emitting gas
\citep{Fabbiano2006}.
The brightest of the discrete sources (L$_x$ $>10^{39}$ ergs s$^{-1}$)
are known as Ultra-Luminous X-ray Sources (ULX), and have attracted
considerable attention in recent years because they have broad-band
X-ray luminosities many times the Eddington limit for a neutron star
or stellar mass black hole \citep[e.g.]{Zezas2002}.  There is a growing concensus that most  ULX can be explained as
high ($\ge$ 10 M$_{\odot}$) stellar mass black holes radiating in excess of the
Eddington limit, possibly with mild beaming  \citep[see]{Roberts2007} and
references therein).  The most massive stellar black holes are thought
to form in low metallicity environments, where the stellar winds of
the progenitor are relatively weak and do not carry off such a large
fraction of the initial mass of the star \citep{Belczynski2004}.
Thus ULX should form preferentially in low metallicity environments.
Observational evidence for this is seen in some ULX  \citep[e.g.]{Soria2005}

\begin{figure*}
\centering
  \includegraphics[angle=0, width=0.8\textwidth]{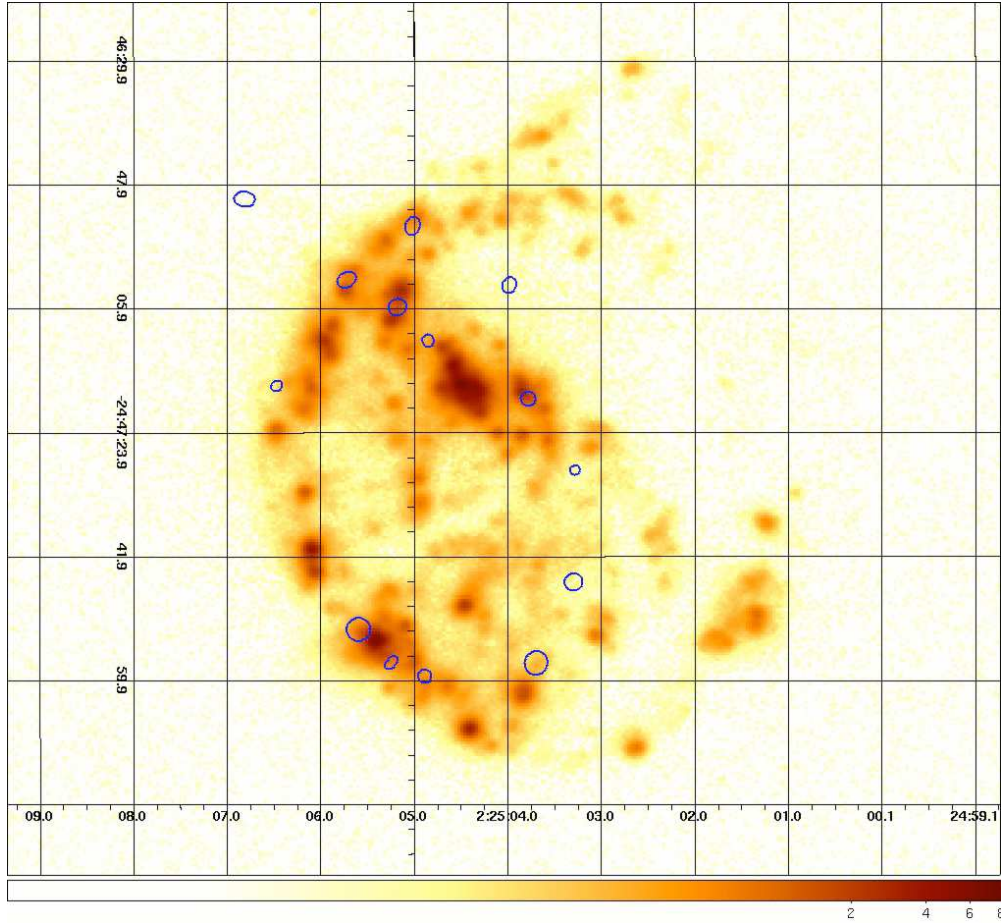}
  \caption{H-$\alpha$ image of NGC 922 with positions of X-ray sources
    marked.  The H-$\alpha$ image is from \cite{Meurer2006}, downloaded
  via the NASA Extragalactic Database.}
  \label{halpha}
\end{figure*}

NGG~922 is a peculiar galaxy with a distinctive C-shaped ring of H$\alpha$
emission (see Figure~\ref{halpha}).  It has been described as a dust-obscured grand design
spiral galaxy \citep{Block2001}.   However, a recent paper by \cite{Wong2006} presents compelling evidence that the morphology could be explained as a result of
an off-center collision between a spiral and its dwarf companion.
This model successfully explains the C-shaped morphology.  The intruder galaxy is identified  to be
a dwarf galaxy approximately 8 arcminutes  away, apparently connected
to NGC~922 by a plume of tidal debris.   The starburst in NGC~922 is still on-going as evidenced by the luminous  H$\alpha$ emission and a population of very young, massive star clusters \citep{Pellerin2010}.   NGC~922 is similar to the
Cartwheel in that both of them are the remnants of drop through
collisions, and both of them have high star formation rates. 
 The Cartwheel galaxy is famous for hosting  12 ULX,  most of which are
 in the star forming ring (\cite{Wolter2004}, see also \cite{Gao2003}).
 This is the largest number of ULX seen in a single system, and has
 led to speculation that the low metallicity of the Cartwheel
 (0.3 \zzero) may optimize the conditions for ULX formation
 \citep{Mapelli2009}.   In contrast, NGC 922 has metallicity near solar
 \citep{Wong2006}.   We are currently engaged in a \chandra\ and \hst\
 study of the X-ray binary population of NGC 922.   In this paper, we
 compare the ULX population of NGC 922 and the Cartwheel.    Our
 primary goal is to determine whether there is an excess of ULX
 (normalized to the star formation rate) in the Cartwheel compared to NGC 922 that can be attributed to the difference in metallicity.   In addition, we use population synthesis models to estimate the expected number of ULX in both systems.    In a future paper we use \hst\ broad-band and H${\alpha}$ images to study the relationship between the X-ray sources and star clusters in NGC 922.   

\section{Data Analysis}
\label{DA}
\begin{figure*}
\centering
  \includegraphics[angle=0, width=0.8\textwidth]{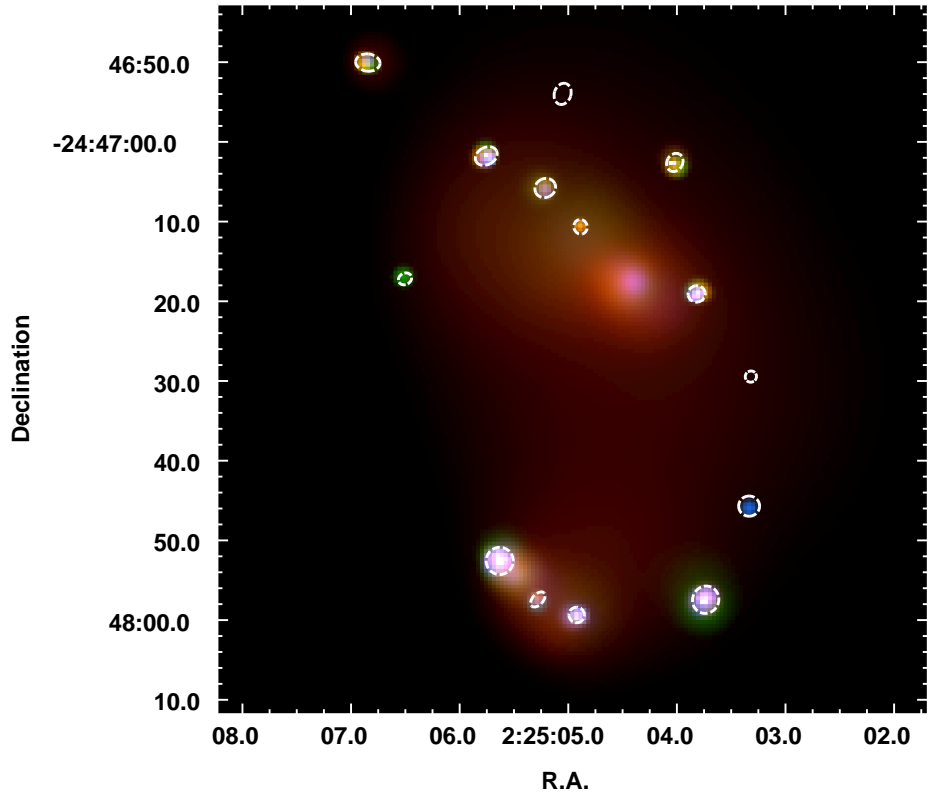}
  \caption{RGB Chandra image of NGC~922. The three bands are \textit{red}: the 0.3\sd1.0\kev; \textit{green}: 1.0\sd2.0; and \textit{blue}: 2.0\sd8.0. An adaptive smoothing algorithm has been applied to the data. The extraction regions for the point sources are defined by the white ellipses.}
  \label{image}
\end{figure*}

Two images of NGC~922 were obtained in 2009.   The first was a 30ks  exposure designed to detect X-ray point sources down to a luminosity of 7$\times10^{38} $ergs s$^{-1}$ and the second  a 10ks exposure, scheduled a few months later, to detect variability in the brightest sources.   Observational details are given in Table~\ref{tab:obs}.    Data were reduced using the software package CIAO (ver. 3.4) with the appropriate calibration products. We created exposure maps using the task \texttt{merge\_all}\footnote{See http://cxc.harvard.edu/ciao/threads/merge\_all.} and we determined the positions of the point sources using the CIAO task \texttt{wavdetect}.\footnote{See http://cxc.harvard.edu/ciao3.0/download/doc/detect\_html\_manual/Manual.html.}     A smoothed X-ray image of NGC~922 is shown in Figure~\ref{image}.   The X-ray point sources are marked.    Counts for the point sources were extracted from regions shown in Figure~\ref{image}.    Background counts for each point source were extracted from an annular region around  each source (these regions are not shown in Figure~\ref{image}).   The counts for each source were corrected for the PSF fraction (Point Spread Function fraction,  i.e.  the fraction of counts outside the extraction region) by using a PSF with energy 1 keV.    The background subtracted, PSF corrected count rate for each  source at both epochs is shown in Table~\ref{data}.   

  We extracted X-ray spectra and standard response files (ARF and RMF)
  for each source.   Fluxes and luminosities for the deep March
  observation were constructed in the 0.3-8 keV band, assuming a
  galactic absorption $N_h$=1.6$\times10^{20}$ cm$^{-2}$.    Using the
  results of \cite{Swartz2004}, we adopt an  intrinsic source spectrum
  with photon index $\Gamma=1.7$  to calculate luminosities.    Fluxes
  were then calculated in XSPEC  by rescaling the normalization of the
  power law model to the observed count rates.    Luminosities were
  calculated assuming a distance of 48 Mpc \citep{Olivares2010}.

\begin{table}
\begin{tabular}{cccccc}
\hline \\
ObsID &  Date  & Exposure Time & Chip & Data Mode & Exposure Mode\\
& & ks & & & \\
\hline\\
10563 & 2009-03-05 & 29.737 & S3 & TIMED EXPOSURE & VFAINT\\
10564 & 2009-10-02 & 10.020 & S3 & TIMED EXPOSURE & VFAINT \\
\hline\\
\end{tabular}
\label{tab:obs}
\caption{Observing configuration for two \chandra\ observations of NGC~922}
\end{table}

\begin{table}
\begin{tabular}{lcccccc}
\hline \\
& & & \multicolumn{2}{c}{Net Count Rate} & F$_X$({\small 0.3-8 keV}) &
L$_X$({\small 0.3-8 keV}) \\
Source & RA & Dec &{\small March 09}&  {\small Oct 09} &10$^{-15}$ ergs cm$^{-2}$ s$^{-1}$ & 10$^{39}$ ergs s$^{-1}$ \\
    &\multicolumn{2}{c}{J2000} &\multicolumn{2}{c}{counts ks$^{-1}$}
    & March 09 & March 09\\
    \hline\\
1 & 02 25 06.84 & -24 46 50.05 & 0.92 & 0.41  &  6.77$\pm$ 1.25 & 1.86  \\
2 & 02 25 06.50  &-24 47 17.19& 0.30  & $<$0.1&  2.49 $\pm$ 0.78 & 0.68 \\
3 & 02 25 05.76 & -24 47 01.78& 1.57 & 2.04 & 14.04 $\pm$ 2.0 &  3.87  \\
4 & 02 25 05.63 & -24 47 52.58 &11.47 & 4.78  & 87.80 $\pm$ 4.64  &  24.20 \\
5 & 02 25 05.28  & -24 47 57.41& 0.49 & 1.06 &4.30$\pm$ 1.05   & 1.19 \\
6 & 02 25 05.21  &-24 47 05.79 & 0.44  & 0.80 &3.74  $\pm$ 1.04 &1.02  \\
7 & 02 25 05.05  & -24 46 53.98 & 0.21 & $<$0.1 &1.89$\pm$  0.79  & 0.52 \\
8 & 02 25 04.92  &-24 47 59.36 & 1.49 & 1.00&11.27 $\pm$ 1.65  &  3.10\\
9  &  02 25 04.89  & -24 47 10.66 &0.44 & 0.29 &3.57 $\pm$ 0.95  &  0.98 \\
10 & 02 25 04.02  &-24 47 02.58 & 1.46 & 1.16 &12.11 $\pm$ 1.77  & 3.33\\
11 & 02 25 03.82 & -24 47 19.08 & 1.68 & 3.28&  14.20 $\pm$1.89  & 3.91  \\
12 & 02 25 03.74 &-24 47 57.45& 4.08 &3.49& 31.49 $\pm$  2.81   &8.67\\
13  & 02 25 03.33 & -24 47 45.70 & 0.42  &0.49&3.23 $\pm$ 0.92  & 0.88  \\
14 & 02 25 03.32  &-24 47 29.45 &0.26  & 0.21 &2.35 $\pm$  0.77  & 0.64 \\
\hline\\
\end{tabular}
\label{data}
\caption{NGC 922 detected X-ray sources}
\end{table}

\section{The X-ray Source Population of NGC 922}
NGC 922 has a population of bright X-ray point sources, including 9 
ULX, defined by L${_x}>10^{39}$ ergs s$^{-1}$, in the 0.3-8.0 keV band
(there are 7 ULX in the 2-10 keV band).  Count rates for
both the March and October observations are shown in Table~\ref{data}.
Many of the sources are variable.    Figure~\ref{halpha} shows a
H$\alpha$ image of NGC 922 with positions of the X-ray sources
superimposed.   The X-ray sources are clearly associated with recent
star formation, and several are coincident with the star forming ring.
The close association of bright X-ray sources with recent star
formation has been observed in many starburst galaxies \citep{Wolter2004, Zezas2007, Rappaport2010}.    It is generally
accepted that the X-ray source population at high luminosities
(L${_x}>10^{37}$ ergs s$^{-1}$) is dominated  by accreting X-ray
binaries \citep{Fabbiano2006}.  

\subsection{The Brightest Source in NGC 922}
\label{x4}

NGC~922~X-4 has an 0.3-8 keV luminosity  L$_x$=2$\times10^{40}$ ergs s$^{-1}$, placing it at the upper end of the X-ray binary luminosity function \citep{Grimm2003, Swartz2004}.    The count rate declined by a factor of 2 in the second observation.    We fit the  0.3-10 keV spectrum of NGC 922 with two models -- a simple power law and a multicolored disk blackbody.   In both cases, a galactic N$_H$ of 1.62$\times10^{20}$ cm$^{-2}$ was assumed and fixed in the fit, and another absorption component intrinsic to the source was allowed to vary.    The resulting fit parameters are shown in Table~\ref{spec}.     The luminosity, spectrum and variability of NGC~922~X-4 are typical of ULX that appear to have hard X-ray spectra at the highest accretion rates   \citep{Gladstone2009,Berghea2008}.     Other examples include IC 342 X-1, NGC 1313 X-1 \citep{Gladstone2009}, NGC 1365 \citep{Soria2009} and  Holmberg IX X-1 \citep{Kaaret2009, Vierdayanti2010}.      This high flux/hard spectrum behavior  is interesting because it appears to be different to what is generally observed in Galactic black hole binaries,  where soft  thermal emission from a disk dominates higher accretion states.    The "high/hard" state as been provisionally called the "ultraluminous" state  \citep{Gladstone2009}.  It may correspond to a super-Eddington accretion state where an out-flowing wind from the inner accretion disk leads to an increased optical depth and lower coronal temperature \citep{Gladstone2009}.

\begin{table}
\centering
\begin{tabular}{ccc}
\hline \\
Parameter&Powerlaw & Multicolored disk \\
\hline\
N$_{H, Gal}$  (cm$^2$) &  1.62$\times10^{20}$ & 1.62$\times10^{20}$ \\
N$_H$ (cm$^2$) & 2.4$^{+1.1}_{-0.9} \times10^{21}$ & 3.4$^{+3.4}_{-9.5} \times10^{20}$\\
&$\gamma$=1.9$^{+0.2}_{-0.3}$ & T$_{in}$(keV)=1.3$\pm$0.3\\
normalization &2.3$\pm$0.95$\times10^{-5}$&1.36$\pm$1.01$\times10^{-3}$\\
\hline\\
$\chi^{2}/DOF$ & 1.15 (20.79/18)& 1.15 (20.7/18)\\
\hline\\
\end{tabular}
\label{spec}
\caption{Best fitting parameters for source X-4.  Models are {\tt phabs*phabs*pow} and {\tt phabs*phabs*diskbb}.   
}
\end{table}

\subsection{Comparison with the Cartwheel}

\begin{figure*}
\centering
  \includegraphics[angle=0, width=0.8\textwidth]{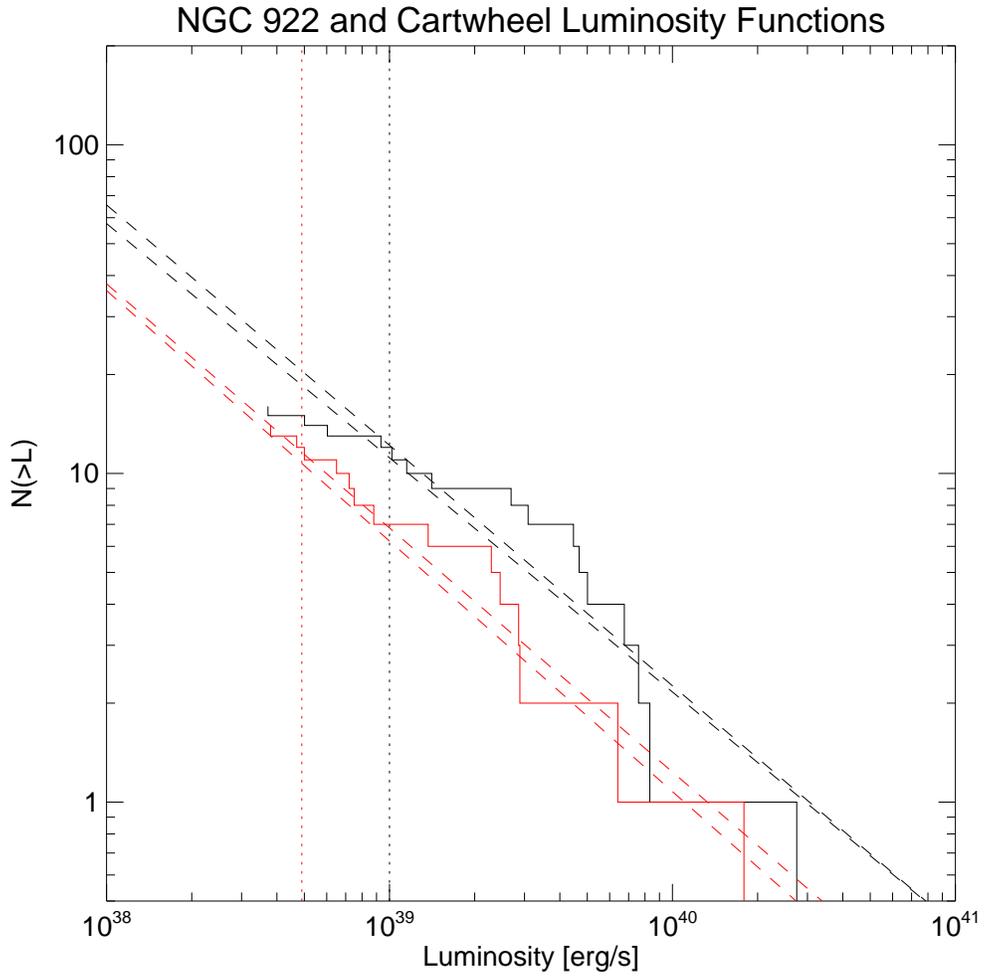}
  \caption{X-ray luminosity function of the Cartwheel galaxy (top,
    black) and NGC~922 (bottom, red).  Two fits are shown for each
    galaxy: the upper fit is to the raw data, and
    the lower fit is corrected for background sources using the
    results of \cite{Giacconi2001}.   The vertical lines show the
    lower limit to the fit for the Cartwheel (black) and NGC~922 (red). }
  \label{lf}
\end{figure*}

Here we compare the ULX population seen in NGC 922 with that of the
Cartwheel.    The Cartwheel has a metallicity in the range Z$\sim
0.18-0.3Z_{\odot}$  in the ring \citep{Fosbury1977, Vorobyov2001}.
This is considerably lower than the metallicity of  NGC 922, which is
$\sim$0.75 Z$_{\odot}$ \citep{Wong2006}.     Figure~\ref{lf} shows the
cumulative luminosity function for sources in NGC 922 and the
Cartwheel.   Luminosities for sources in the Cartwheel were taken from
\cite{Wolter2004}.   In order to check that there were no systematic
differences in fluxes derived using our method (described in \ref{DA})
and that used by \cite{Wolter2004} we repeated our analysis using the
archival Cartwheel data.  We derived fluxes essentially identical to
\cite{Wolter2004}.  We note that luminosites for X-ray sources in the
Cartwheel are calculated  in the 2-10 keV band \citep{Wolter2004}.
The X-ray luminosities of  sources in NGC~922 have been plotted in the
2-10 keV band in order to allow for direct comparison with the
Cartwheel.  This results in slightly lower luminosities than given in
Table~\ref{data}.   There are 7 ULX in NGC~922 in the 2-10 keV band.

  We fit a single power law to the unbinned, differential,  XLFs using
  a maximum likelihood statistic \citep{Crawford1970}.   The
  goodness-of-fit (GOF) estimate is performed by simulating a
  luminosity distribution with the best-fit slope. One million
  iterations were performed. If the data are well fitted by a single
  power law, the GOF statistic approaches 1.0. The contribution of
  cosmic background sources to the XLF was evaluated using the log
  N-log S curves of \cite{Giacconi2001}.  Details of the fitting method are given in \cite{Kilgard2005}.  Table~\ref{xlf_fits} gives the results. 

Table~\ref{xlf_fits} shows that the XLF slopes of high luminosity
X-ray sources in the Cartwheel and NGC 922 are identical within the
uncertainties.   The slopes are consistent with the ``Universal'' High
Mass X-ray Binary XLF described by \cite{Grimm2003}.   Assuming that
the star formation rates of the
Cartwheel and NGC 922 are 18  and 8 M$_{\odot}$/yr respectively
\citep{Mayya2005,Wong2006},  we find that the number of ULX sources
relative to the star formation rate is 0.94$\pm$0.3  ULX per solar
mass of star formation for NGC 922, and 0.67$\pm$0.2 ULX per solar
mass of star formation for the Cartwheel.   Here we use the number of
ULX in the 2-10 keV band. These ratios are dependant on the assumed
star formation rates.  The star formation rates quoted for NGC 922 and the Cartwheel were derived
using extinction-corrected H$\alpha$ luminosities.   There is no
evidence for widespread obscured star formation in either galaxy, so
these values are likely to be close to the correct values.  
 We conclude there is no evidence for an excess of high luminosity sources
(normalized to the star formation rate) in the lower-metallicity
Cartwheel unless the SFR of either galaxy is wrong by a factor of 2 or
more.

\begin{table}
\centering
\begin{tabular}{lccccc}
\hline \\
Galaxy &$\gamma$ &L$_{min}$  & $GOF$ & $N_{fit}$&N$_{39}$\\
& & ergs s$^{-1}$ & & &  \\
\hline\\
Cartwheel & 0.71$\pm$0.120 & 1$\times 10^{39}$ &   0.81 &12.0 &11.1\\
NGC 922 & 0.76$\pm$0.22& 5$\times 10^{38}$ & 0.71 & 14 &7.5 \\
 \hline\\
\end{tabular}
\label{xlf_fits}
\caption{Power law fits to the  \xlf.  $\gamma$ is the fitted slope to
  the \xlf\ correcting for background sources.   L$_{min}$ is the
  minimum luminosity included in the fit.   GOF is the
  goodness-of-fit parameter, $N_{fit}$ the number of sources
  included in the fit and N$_{39}$ the predicted number of ULX  (i.e. sources
  with $L_x > 10^{39}$ ergs s$^{-1}$)   } 
\end{table}

\section{Population Synthesis Models}



The similar ULX/SFR observed in the Cartwheel and NGC 922 is, in fact, expected in population synthesis models of ULX formation in low metallicity environments. \citet{Linden2010} studied the mechanisms for high mass X-Ray binary formation in starburst environments, finding the ULX population to be dominated by systems moving through pathways powered by strong wind accretion from supergiant donors onto massive black holes formed in failed supernovae at both Z~=~0.75~Z$_\odot$ and  Z~=~0.28~Z$_\odot$.

Utilizing simulations from the StarTrack code~\citep{Belczynski2008} simulations using a SFR of
18~M$_\odot$~yr$^{-1}$ for the Cartwheel and 8~M$_\odot$~yr$^{-1}$ for
NGC 922~\footnote{The continuous star formation period is simulated to
  last for 100~Myr. However, a negligible number of ULX are produced
  after 20~Myr, rendering the exact end of the simulation unimportant.}
we calculate a population of 103 ULX for NGC 922 and 238 ULX for the
Cartwheel. We note two results: (1) our default simulations
overpredict the absolute number of observed systems by approximately
an order of magnitude, and (2) our models similarly predict no
metallicity dependence in the ULX/SFR ratio in the regime between
0.75~Z$_\odot$ and 0.3~Z$_\odot$ investigated in this study.   

Predictions regarding the absolute number of bright X-ray Binaries are particularly difficult for population synthesis models. We note several obvious parameter space choices which would allow a match between our simulations and observations. Because the supergiant donors have extremely large radii, the ULX formed through this pathway must be weakly bound - implying that small natal kicks imparted to massive BHs (e.g. a Maxwellian distribution with dispersion velocity 26.5~km~s$^{-1}$) can disrupt as many as 90\% of ULX progenitors~\citep{Linden2010}. Additionally, our definition of ULX assumes an average X-Ray luminosity of 1~x~10$^{39}$~erg~s$^{-1}$, however eccentricities may make these systems dimmer for a substantial portion of their orbit. Another plausible explanation concerns the treatment of super-Eddington accretion. In our default simulations we allow black hole accretors to power ULX at up to ten times the Eddington limit, so long as the donor winds are able to supply enough material to the black hole accretor. Our models do not describe the specific mechanisms which allow this. Several mechanisms involving dynamic instabilities have been postulated - and include leaky-disk \citep{Begelman2002} or beaming \citep{King2001} models which may imply a duty cycle at the 10\% level.

The ratio of ULX formation rates between NGC 922 and the Cartwheel
places more meaningful constraints on the models than absolute
numbers,  as there are fewer parameters which can fine tune the
metallicity dependence. In this case, we find models using the default
assumptions in \citet{Linden2010} to correctly predict insignificant
metallicity dependence in the regime from Z~=~0.75~Z$_\odot$ to
Z~=~0.18~Z$_\odot$. We emphasize that this relation falls naturally out
of our simulations, as the ULX population at both metallicities is
created through the same evolutionary channels. We note that if the
metallicity of the Cartwheel is significantly lower than assumed here,
then this assumption would no longer hold, and the Cartwheel would
begin to be dominated by a second population powered by Roche Lobe
overflow of main sequence stars onto fallback BH accretors (this model
has been suggested as a mechanism to explain the properties of the
brightest source in the Cartwheel \citep{Pizzolato2010}).  Thus the
consistency of the ULX/SFR ratio  provides some preference for a
higher modeled metallicity within the Cartwheel. Guided by this
consistency in the ULX/SFR in the observed metallicity range, we note
several other robust results. First, the supergiant donors which power
ULX activity die off quickly after star formation, and we find
$\sim$95\% of luminous ULXs to be younger than 20~Myr. Secondly, the
direct collapse pathway imparts almost no systematic velocity to the
resultant X-ray binaries , and thus ULX are constrained to lie very close to recent star formation regions. However, these systems may gain significant spatial velocities if they are formed within dense stellar environments where three body interactions become non-negligible \citep{Mapelli2011}.  Finally, we caution that the significant overabundance of ULX sources
predicted by StarTrack models implies an additional physical mechanism
which must eliminate a large number of possible ULX systems. We are
unable, in this study, to constrain this mechanism from introducing a
metallicity dependence, which would create a mismatch between
observations and our simulated population.

\section{Summary and Conclusions}

NGC 922 is a drop-through ring galaxy \citep{Wong2006} with an
expanding ring of star formation, similar in many respects to the
Cartwheel galaxy.   New Chandra observations reveal  a population of
bright X-ray point sources, including nine  ULX, defined by
L${_x}>10^{39}$ ergs s$^{-1}$ in the 0.3-8.0 keV band.    The brightest source  has an 0.3-8 keV luminosity  L$_x$=2$\times10^{40}$ ergs s$^{-1}$, placing it at the upper end of the X-ray binary luminosity function \citep{Grimm2003, Swartz2004}.    The spectrum of this source is hard, and similar to other binaries in an "ultraluminous" state  \citep{Gladstone2009,Berghea2008}. 

We compare the NGC~922 X-ray source population  with that of the
Cartwheel.  These two systems are similar in many respects (both
drop-through ring galaxies) but the Cartwheel has a lower metallicity
than NGC 922.   We find that the number of ULX in NGC 922 and the Cartwheel scales with the star formation rate:  there is no evidence for an excess of sources in the Cartwheel.        

Population synthesis  models using \startrack\ suggest that the ULX population in both NGC 922 and the Cartwheel is dominated by systems with strong wind accretion from supergiant donors onto  direct-collapse BHs.       The default  models over-predict the number of observed ULX in both galaxies.   There are several simplifying assumptions in the simulations which could account for the discrepancy.   For  example, the models assume that direct collapse events impart no natal kick to the resultant black hole.  However, even small kicks could separate the widely spaced progenitors before the ULX phase begins,  hence reducing the number of ULX.   The simulations correctly predict the ratio of the number of ULX in NGC 922 and the Cartwheel.   We emphasize that this relation falls naturally out of the \cite{Linden2010} simulations, as the ULX population at both metallicities is created through the same evolutionary channels.   We note that the  \citet{Linden2010} simulations predict   a substantially different ULX population below  Z $<$0.15~Z$_\odot$. Thus it would appear that the metallicity of the Cartwheel is not low enough to see a difference in the ULX population compared to NGC 922.    We are currently working on a project to search for ULX in  Extremely Metal Poor Galaxies (XMPG).  These are the most metal poor galaxies known, and a logical place to find ULX if they favor metal poor systems.     We will report on these findings in a future paper.

\section{Acknowledgements}

We thank the referee for comments which improved this paper.  Support for this work was provided by the National Aeronautics and
Space Administration through Chandra Award Number GO9-0097A issued by the
Chandra X-ray Observatory Center, which is operated by the Smithsonian
Astrophysical Observatory for and on behalf of the National
Aeronautics Space Administration under contract NAS8-03060.   This research has made use of the NASA/IPAC Extragalactic Database (NED) which is operated by the Jet Propulsion Laboratory, California Institute of Technology, under contract with the National Aeronautics and Space Administration. GT and AW acknowledge partial financial contribution from the
 ASI-INAF agreement  I/009/10/0.

\label{lastpage}

\end{document}